\documentclass[twocolumn,preprint,]{aastex63}

\usepackage{amsmath,amstext}
\usepackage[T1]{fontenc}
\usepackage{apjfonts} 
\usepackage{graphicx}
\usepackage{natbib}
\usepackage{longtable}
\usepackage{CJK}
\usepackage{microtype,longtable,verbatim,float,booktabs,graphicx}

\newcommand{\HavNII}{\hbox{({\rm H}$\alpha$+{\rm [N}\kern 0.1em{\sc II}{\rm ]})}}

\newcommand{\HnFeV}{\hbox{{\rm H}9+{\rm [Fe}\kern 0.1em{\sc V}{\rm ]}}}

\newcommand{\HeH}{\hbox{{\rm H}8+{\rm He}\kern 0.1em{\sc I}{\rm }}}

\newcommand{\HeII}{\hbox{{\rm He}\kern 0.1em{\sc II}}}

\newcommand{\HeI}{\hbox{{\rm He}\kern 0.1em{\sc I}}}

\newcommand{\HII}{\hbox{{\rm H}\kern 0.1em{\sc II}}}

\begin{document}

\title{\large \bf MEGA and SMILES Find Fewer Dusty Galaxies than Expected at Cosmic Noon}
\begin{CJK}{UTF8}{gbsn}

\author[0000-0001-8534-7502]{Bren E. Backhaus}
\affil{Department of Physics and Astronomy, University of Kansas, Lawrence, KS 66045, USA}

\author[0000-0002-5537-8110]{Allison Kirkpatrick}
\affiliation{Department of Physics and Astronomy, University of Kansas, Lawrence, KS 66045, USA}

\author[0000-0002-6292-4589]{Kurt Hamblin}
\affiliation{Department of Physics and Astronomy, University of Kansas, Lawrence, KS 66045, USA}

\author[0000-0001-5749-5452]{Kaila Ronayne} 
\affiliation{Department of Physics and Astronomy, Texas A\&M University, College Station, TX, 77843-4242 USA} 
\affiliation{George P.\ and Cynthia Woods Mitchell Institute for Fundamental Physics and Astronomy, Texas A\&M University, College Station, TX, 77843-4242 USA}

\author[0000-0002-9921-9218]{Micaela Bagley}
\affiliation{Astrophysics Science Division, NASA Goddard Space Flight Center, 8800 Greenbelt Rd, Greenbelt, MD 20771, USA}

\author[0000-0001-8519-1130]{Steven L. Finkelstein}
\affiliation{Department of Astronomy, The University of Texas at Austin, Austin, TX, USA}
\affiliation{Cosmic Frontier Center, The University of Texas at Austin, Austin, TX, USA}

\author[0000-0002-8360-3880]{Dale D. Kocevski}
\affiliation{Department of Physics and Astronomy, Colby College, Waterville, ME 04901, USA}

\author[0000-0001-9187-3605]{Jeyhan S. Kartaltepe}
\affiliation{Laboratory for Multiwavelength Astrophysics, School of Physics and Astronomy, Rochester Institute of Technology, 84 Lomb Memorial Drive, Rochester, NY 14623, USA}

\author[0000-0002-6610-2048]{Anton M. Koekemoer} 
\affiliation{Space Telescope Science Institute, 3700 San Martin Drive, Baltimore, MD 21218, USA} 

\author[0000-0003-3216-7190]{Erini Lambrides}
\altaffiliation{NPP Fellow}
\affiliation{NASA-Goddard Space Flight Center, Code 662, Greenbelt, MD, 20771, USA}

\author[0000-0001-7503-8482]{Casey Papovich} 
\affiliation{Department of Physics and Astronomy, Texas A\&M University, College Station, TX, 77843-4242 USA} 
\affiliation{George P.\ and Cynthia Woods Mitchell Institute for Fundamental Physics and Astronomy, Texas A\&M University, College Station, TX, 77843-4242 USA} 

\author[0000-0001-5930-0532]{Gregory Troiani}
\affil{Department of Physics and Astronomy, University of Kansas, Lawrence, KS 66045, USA}

\author[0000-0001-8835-7722]{Guang Yang(杨光)}%
\affiliation{Nanjing Institute of Astronomical Optics \& Technology, Chinese Academy of Sciences, Nanjing 210042, China}
\affiliation{CAS Key Laboratory of Astronomical Optics \& Technology, Nanjing Institute of Astronomical Optics \& Technology, Nanjing 210042, China}

\begin{abstract}

We present infrared (IR) luminsosity functions (LFs) and resulting star formation rate densities using the JWST Mid-infrared Instrument (MIRI) observations from the MIRI EGS Galaxy and AGN (MEGA) survey and Systematic MIRI Legacy Extragalactic Survey (SMILES). JWST allows us to perform a robust analysis on the faint end of the IR LF beyond the local universe. 
We directly measure the 7.7$\mu$m polycyclic aromatic hydrocarbon (PAH) feature using either F1000W, F1500W, or F2100W photometry.
This results in a sample of 634 galaxies across the two surveys covering an area of 105 arcmin$^2$ ($\sim$70 in the EGS and $\sim35$ in the GOODS-S/HUDF fields) and spanning $0.2<z<2$. We convert the 7.7$\mu$m PAH luminosity to total IR luminosity, resulting in LFs that are two orders of magnitude fainter than previous studies. 
In contrast to previous extrapolations based on shallower observations, we find a strong flattening in the faint end of the LF with an average slope of $\alpha\sim0.147$. This indicates that less luminous galaxies do not have as much dust obscured star formation as predicted. 
We measure the star formation rate density (SFRD) by integrating our new IR LFs and find a slightly lower SFRD in all redshift bins than previous studies made with ALMA, Herschel, and Spitzer. We also measure the contribution to the SFRD as a function of luminosity and confirm that LIRGs and ULIRGs remain the dominant contributors to the dust-obscured star formation at $z\sim1-2$.

\end{abstract}

  

\section{Introduction}

Understanding galaxy luminosity distributions and star formation history is vital for probing how galaxies formed and evolved. The luminosity function (LF) is a measure of how galaxies are distributed in cosmic volumes as a function of their brightness. Building LFs for discreet redshift bins can reveal a clear picture of galaxy evolution, as luminosity is tied directly to star formation.
The LF can then be used to determine the star formation rate density (SFRD) in the universe over cosmic time.

In the ultraviolet (UV), LFs measure the unobscured star formation rate, as UV light emanates directly from young stars.
Thanks to the sensitivity of JWST, the UV LF has now been measured out to $z>9$, and possibly even up to $z\sim25$ \cite[e.g.][]{Oesch2018,Bouwens2023,Donnan2024,Perez2025}
The UV LF reveals a strong evolution in the number density, luminosity distribution, and faint-end slope as look back time increases. 
 At high redshift ($z>6$), the number density decreases and the faint-end slope steepens ($\alpha<-2$), indicating that low-luminosity galaxies dominate the population and likely drive cosmic reionization \citep[e.g.][]{Bouwens2015,Finkelstein2015}. Observations with JWST have found an abundance of UV-bright galaxies, implying more elevated star formation efficiencies and  minimal dust attenuation \citep[e.g.][]{Fink2023,Donnan2024,Franco2025}.
Dust attenuation can diminish UV emission, causing UV-selected samples to miss a significant (up to 50\%) fraction of star formation \citep[e.g.][]{Puget1996,Hauser2001,Popescu2011,Viaene2016,Hickox2018}. Therefore, in order to have a complete picture of star formation through cosmic time, we must account for dust obscured galaxies \citep{Genzel2000}.
This can be done by using infrared (IR) measurements which arise from dust primarily heated by young stars, allowing us to probe populations that are underrepresented or absent in UV-based studies. 

IR LFs have been studied extensively over the past two decades using observations from Atacama Large Millimetre/submillimetra Array (ALMA) \citep[e.g.][]{Koprowski2017,Gruppioni2020}, the \textit{Herschel Space Observatory} \citep[e.g.][]{Gruppioni2013,Magnelli2013}, and the \textit{Spitzer Space Telescope} \citep[e.g.][]{LeFloch2005}. Generally, these works fix the slopes for the bright and faint end and find a decrease in characteristic volume density ($\Phi^\ast$) and increase in characteristic IR luminosity ($L^\ast$). 
From population studies with Herschel and Spitzer, we learned that the cosmic star formation rate density was dominated by obscured star formation at its peak ($z\sim1-2$), and luminous IR galaxies (LIRGs; $L_{\rm IR}=10^{11}-10^{12}\,L_\odot$) and ultra luminous IR galaxies (ULIRGS; $L_{\rm IR}>10^{12}\,L_\odot$) were the main contributors \citep[e.g.][]{Caputi2007,Murphy2011}. 
However, due to the sensitivity limitations of the previous generation of IR telescopes, these works were unable to measure galaxies with fainter IR luminosities even at moderate redshifts ($1<z<2$).
As a result, the faint end of the IR luminosity has been extrapolated based on various assumptions and hence is highly uncertain. Predictions universally agree that it has a slope indicating an increasing number of galaxies as luminosity decreases, similar to the behavior of the UV LF \citep[e.g.][]{Bouwens2015,Finkelstein2015,Bouwens2023,Franco2025}.

With the launch of the Mid-InfraRed Instrument (MIRI) instrument \citep{Rieke2015} on board the James Webb Space Telescope (JWST), observations can fill in this crucial region and constrain, for the first time, the faint end of the IR LF. MIRI is superior to previous instruments aboard Spitzer due to its lower instrumental noise, thermal noise, larger field of view, sensitivity to warm dust, better angular resolution, and multi-band mid-IR photometry. With this improved design, JWST/MIRI provides the ability to detect and measure polycyclic aromatic hydrocarbons (PAHs) for galaxies in the local universe and beyond. PAHs are complex organic molecules that emit strongly when excited by UV radiation from young stars, making them an excellent way to measure dust obscured star formation. 
Additionally, there is a strong correlation between the 6.2\,$\mu$m and 7.7\,$\mu$m PAH complexes and the IR luminosity ($L_{IR}$, measured from $8-1000\,\mu$m), which traces the star formation rate over the past 100 million years \citep[e.g.][]{Houck2007,Shipley2016,ronayne2024,Shivaei2024}.


In this work, we use JWST/MIRI observations from the Systematic Mid-infrared Instrument Legacy Extragalactic Survey (SMILES) and the MIRI EGS Galaxy and AGN (MEGA) survey to obtain monochromatic and IR ($L_{IR}$) LFs out to $z\sim2$.
In Section \ref{Data}, we discuss the observational set up and data selection of our sample. In Section \ref{Method}, we outline how we calculated and measured the LFs. In Section \ref{Mono_LF}, we discuss the monochromatic LFs while in Section \ref{Eq:LF}, we discuss the resulting IR LFs in discrete redshift bins. In Section \ref{SFRD}, we calculate the star formation rate density by integrating the IR LFs in each redshift bin. Finally in Section \ref{Summary}, we summarize our findings and results. In this work, we assume a $\Lambda$CDM cosmology with $\Omega_{M}=0.3$, $\Omega_{\Lambda}=0.7$, and $H_{0}=70\ {\rm km\ s^{-1}\ Mpc^{-1}}$.

\section{Data Selection and Method}

\subsection{Sample}\label{Data}

We compiled galaxy samples from the MEGA and SMILES survey. The MEGA\footnote{https://kirkpatrick.ku.edu/MEGA/} survey (PI: Kirkpatrick; \cite{backhaus2025}) covers $\sim$70 arcmin$^2$ of the Extended Groth Strip \citep[EGS][]{Davis2007}. It has a 5$\sigma$ depth of 0.41 $\mu$m, 1.26$\mu$m, and 4.10 $\mu$m in the F1000W, F1500W, and F2100W respectively.
The SMILES\footnote{https://archive.stsci.edu/hlsp/smiles} survey (PI: Rieke, \cite{Rieke2024}) covers $\sim$35 arcmin$^2$ of the GOODS-S/HUDF. It has a 5$\sigma$ depth of 0.39 $\mu$m, 0.75$\mu$m, and 2.80 $\mu$m in the F1000W, F1500W, and F2100W, respectively \citep{Alberts2024}.
To obtain redshifts for the galaxies, we combine the source catalog from MEGA \citep{backhaus2025} with the UNICORN project NIRCam photometric catalog \cite[Finkelstein et al. 2026, in prep]{Bagley2023}, and we combine SMILES \citep{Alberts2024} with CANDELS HST catalog \citep{Kodra2023}. Both catalogs include photometric redshift measurements.

\begin{deluxetable}{|c|c|c|c|c|c|c|}[ht!]
\tablecaption{Sample Sizes}
\label{tbl:numbers}
\tablehead{\multicolumn{3}{|c|}{Monochromatic} & \multicolumn{3}{|c|}{$L_{IR}$} &  \\
\hline
 Filter & $N_{gal}$ &$N_{MEGA}$ & $N_{SMILES}$ & $N_{gal}$ &$N_{MEGA}$  & $N_{SMILES}$}
\startdata
F770W & 4051 & 1871 & 2180 & ... & ... & ...\\
    \hline
    F1000W & 3301 & 1944 & 1357 & 136 & 99 & 37\\
    \hline
    F1500W & 2530 & 1366 & 1164 & 245 & 109 & 136 \\
    \hline
    F2100W & 2003 &  1141 & 862 & 253 & 145 & 108\\
    \hline
\enddata
\end{deluxetable}

To be included in the parent sample, from which we construct monochromatic LFs, we require that a source have a flux above the 5$\sigma$ image depths for MEGA and SMILES and have a SNR $>3$. This results in 4150, 3301, 2530, and 2003 galaxies in the F770W, F1000W, F1500W, and F2100W filters, respectively (Table \ref{tbl:numbers}).

To be included in our final sample used to create the IR LFs, a source must have a redshift such that the $7.7\,\mu$m PAH feature falls in one of the photometric filters. This corresponds to the following redshift ranges: $0.24< z< 0.36$ (F1000W), $0.85 < z < 1.1$ (F1500W), and $1.52 < z < 1.90$ (F2100W).
This results in 136 galaxies in F1000W, 245 galaxies in F1500W, and 253
galaxies in F2100W for a total of 634 galaxies being used in the final sample.

\begin{figure}[b!]
\centering
\includegraphics[width=\linewidth]{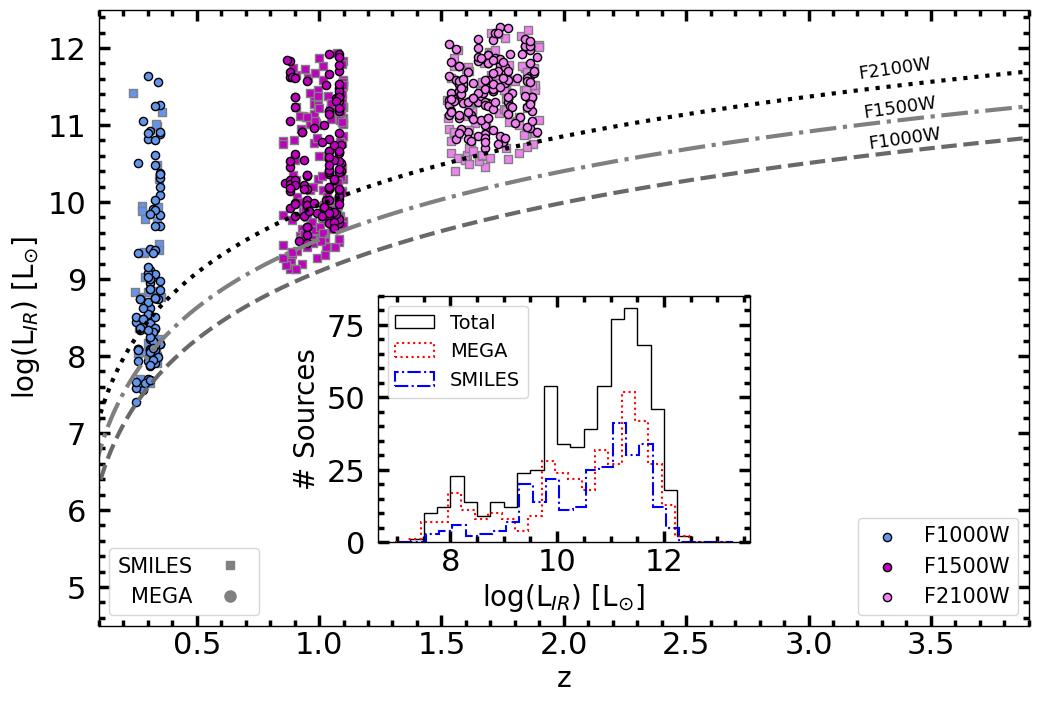}
\caption{$L_{\rm IR}$ plotted against redshift for the MEGA (circles) and SMILES (squares) observations. The lines represent the limiting MEGA $L_{\rm IR}$ created by the 5$\sigma$ flux detection limits. The SMILES limits are similar to MEGA but slightly deeper. The inset plot shows the $L_{\rm IR}$ distribution of the total initial sample (black line) and each of the two surveys (red and blue lines).
\label{fig:L_evo} }
\end{figure}

The redshift and $L_{\rm IR}$ distribution of the final sample is shown in Figure \ref{fig:L_evo}, along with the MEGA survey's limiting $L_{\rm IR}$ based on the 5$\sigma$ depth in each filter. Note that the SMILES galaxies go slightly below this line due to the slightly deeper 5$\sigma$ depths. The inset shows the number of sources at a given IR luminosity for the total sample as well as the individual surveys. The requirement to have the the 7.7$\mu$m PAH feature land in the MIRI filters created three distinct redshift bins; we further split the highest redshift bins created from the F1500W and F2100W observations in half to have five redshift bins in total with approximately the same number of galaxies. 

\subsection{$L_{7.7PAH}$ and $L_{\rm IR}$ Calculations}\label{Method}

We estimate the 7.7$\mu$m PAH luminosity from the filter photometry using a conversion factor following the method in \citet{Kirkpatrick2023}. The conversion factor for each filter was constructed from the 5MUSES \citep{Wu2010} subsample presented in \cite{Kirkpatrick2014} which span $10\le \log L_{\rm IR} [L_\odot]\le 11.7$ and $0.04<z<0.24$. We use this sample due to it including lower luminosity galaxies and its available Spitzer/IRS spectra.
We select the 12 completely star-forming galaxies that span a range of PAH equivalent widths, to find the average strength of the 7.7\,$\mu$m PAH feature. The 7.7$\mu$m PAH feature is constrained between 7.2 and 8.2 microns, with the continuum slope fitted using the wavelengths on either side of the feature. The continuum slope is subtracted, and we integrate the remaining line flux to arrive at the 7.7\,$\mu$m luminosity. We then average $L_{7.7PAH}$ from all 12 galaxies, in order to avoid biasing our calculations by using a single source with strong PAH emission. Using the transmission curve of the F1000W, F1500W, or F2100W filter, we create synthetic photometry for each 5MUSES galaxy at the precise redshift of every source in our sample. We also average the photometry for all 12 galaxies.
We calculate the conversion factor as the ratio of the average $L_{7.7PAH}$ and average synthetic photometry. Finally, we apply these conversion factors to the MEGA + SMILES sample. For example, for a MEGA galaxy where the PAH feature falls in the F1000W filter, we would calculate:

\begin{equation}
L_{7.7PAH}^{\text{MEGA}} = \frac{\overline{L_{7.7PAH}^{\rm 5MUSES}}}{\overline{F_{1000W}^{\rm 5MUSES}}}\times F_{1000W}^{\rm MEGA}
\end{equation}

We obtain the IR luminosity through the conversion of $L_{7.7PAH}$. We use two separate conversions depending on a galaxy's $L_{7.7PAH}$.  We  created a $7.7\,\mu$m LF to find the knee ($L_{7.7PAH}^\ast$) in each redshift bin. Below the knee, we use the conversion from \cite{Shivaei2024}:
\begin{equation} \label{Eq:PAH_IR}
  \log(L_{IR}) = \frac{\log(L_{7.7PAH}) -0.31}{0.83}\
\end{equation} 
This conversion is derived using 76 purely star-forming galaxies with $9.5\le \log L_{\rm IR} [L_\odot]\le13.0 $ covering $0<z<3$ amassed from new ALMA observations and previously published samples, including the 5MUSES sample from \citet{Kirkpatrick2014}.

For our galaxies above the knee, we use the relation in \cite{Houck2007}, which is derived from Spitzer/IRS spectroscopy (to directly measure $L_{7.7PAH}$, MIPS photometry, and IRAS far-IR photometry for 50 galaxies at $z<0.6$ with $9.0 \le \log L_{\rm IR} [L_\odot]\le 12.5$:

\begin{equation} \label{Eq:PAH_IR}
  \log(L_{\rm IR}) = \log(L_{7.7PAH})+0.78 \
\end{equation} 
where $L_{\rm IR}$ and $L_{7.7PAH}$ are in units of $L_\odot$.

\begin{figure}[h!]
\centering
\includegraphics[width=\linewidth]{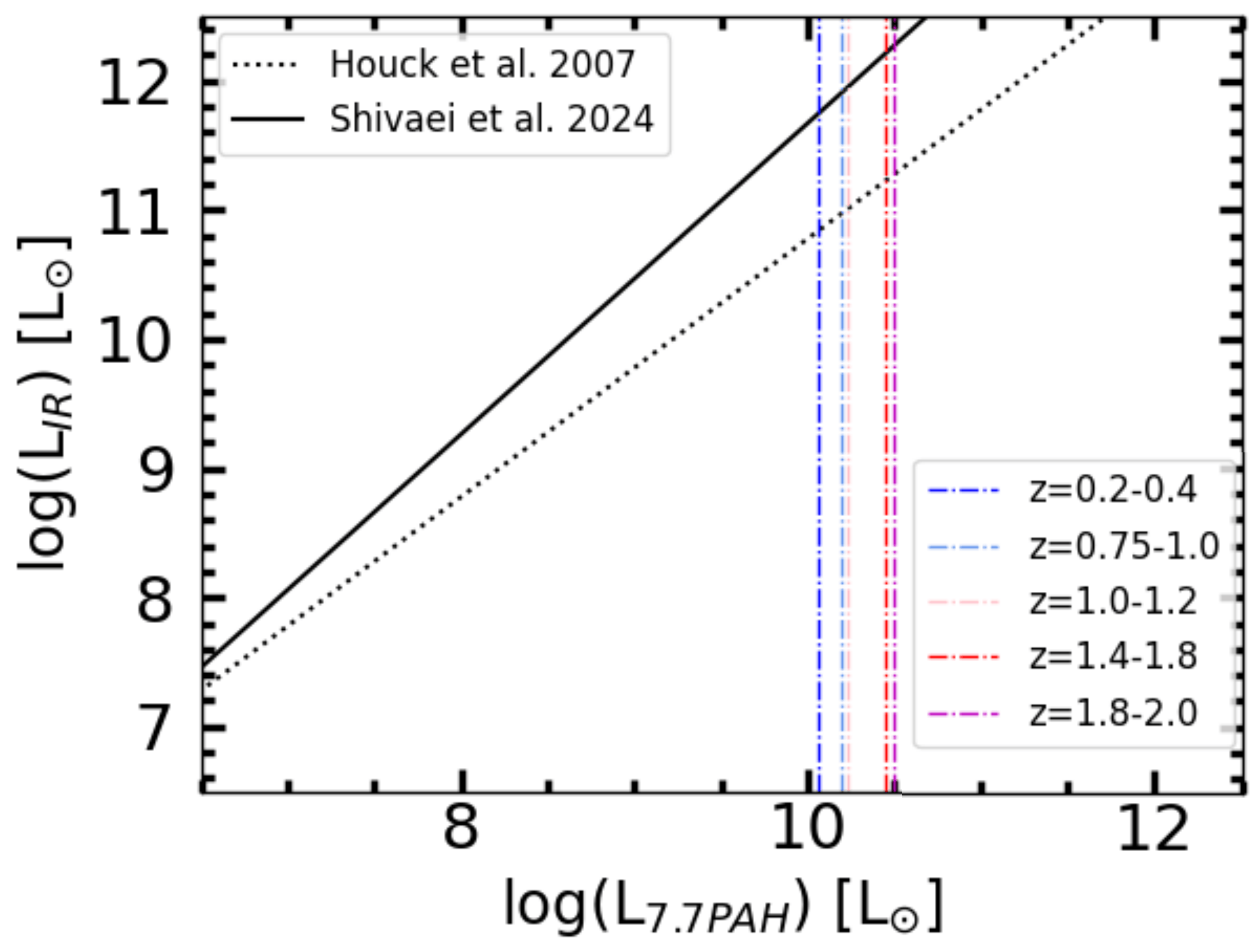}
\caption{Difference in the resulting $L_{\rm IR}$ from the \cite{Shivaei2024} and \cite{Houck2007} conversions at a given $L_{7.7PAH}$. By $L_{7.7PAH}$=10, there is a 1.1\,dex difference. The colored lines show the $L_{7.7PAH}$ knee in each redshift bin. Below the knee,  we use the \cite{Shivaei2024} conversion, and above the knee, we use \cite{Houck2007}.
\label{fig:L_meth_dif} }
\end{figure}

\begin{figure*}[t!]
\centering
\includegraphics[width=\linewidth]{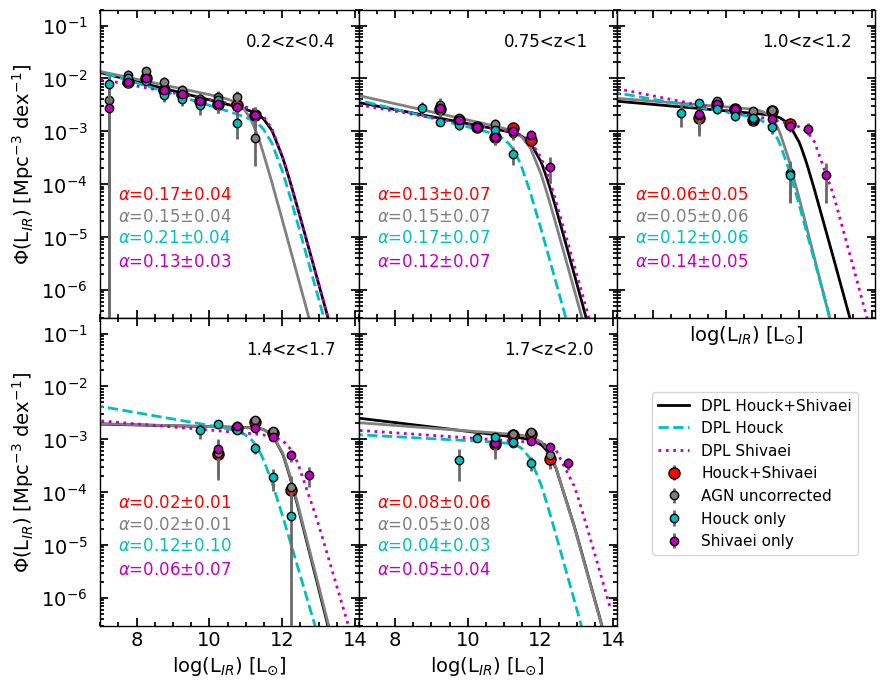}
\caption{Integrated IR LFs (methodology described in \S4) for five redshift ranges for different conversion methods between the 7.7$\mu$m PAH luminosity and $L_{\rm IR}$ as well as the AGN uncorrected method of using both \cite{Shivaei2024} and \cite{Houck2007}. The final fit lines for all methods are shown, with the slope of the faint end annotated in the same color as the method it is associated with. The values for these alternate LFs can be found in Appendix A.
\label{fig:LF_methods} }
\end{figure*}

Figure \ref{fig:L_meth_dif} highlights how these two conversions diverge at higher $L_{7.7PAH}$ values. The full methodology for calculating the functional form of the the LFs will be described in Section \ref{sec:LF}. Initially, we used only the \cite{Shivaei2024} relation for all luminosity bins as it is a more recent work; however, as shown in Figure \ref{fig:LF_methods} (purple points), we found it leads to a significant deviation in the higher luminosity bins at $z>1$ compared to previous observations \citep{LeFloch2005,magn11,Casey2012,Gruppioni2013,Magnelli2013,Gruppioni2020,Traina2024,Fujimoto2023}. Using the \cite{Houck2007} conversion in the high luminosity bins allow us to do a direct comparison with previous literature results.

To validate our method for estimating $L_{\rm IR}$ from the 7.7 $\mu$m PAH emission, we compare our values to those derived from far-infrared observations. To do this we create a subsample with Herschel/SPIRE 250\,$\mu$m observations from the Herschel Multi-tiered Extragalactic Survey \citep[HerMES][]{Oliver2012}. For each galaxy, we derived $L_{\rm IR}$ using our adopted methodology and compared the results to independent $L_{\rm IR}$ estimates obtained from deblended Herschel/SPIRE F250$\mu$m photometry through established far-infrared spectral energy distribution fitting procedures.We find good agreement between the measurements with a median offset of $\Delta log(L_{\rm IR})\sim0.11$. This comparison demonstrates that our approach provides robust $L_{\rm IR}$ estimates consistent with those derived from direct far-infrared observations.

As a result of JWST's ability to provide robust measurements of the 7.7$\mu$m emission in galaxies with significantly lower IR luminosities, PAH-based estimates of $L_{\rm IR}$ can be obtained for large samples of star-forming galaxies which are undetected in existing far-IR observations as these are biased towards the brightest galaxies.
Without a sensitive far-IR telescope, we are forced to extrapolate far-IR emission from mid-IR data. While this can be done through SED fitting, that introduces a strong dependence on choice of templates, initial assumptions, and fitting routines. Therefore, empirically-derived relations between the 7.7$\mu$m PAH feature and $L_{\rm IR}$ provides a practical and efficient proxy for JWST samples.


\subsection{Calculating Co-Moving Volume}

We derive the LFs using the 1/$V_{max}$ method as it does not make any assumptions about the shape of the fuction \citep{Schmidt1968}. 
The LF for each redshift and luminosity bin is:
\begin{equation} \label{Eq:obsLF}
  \Phi(L,z) = \frac{1}{\Delta L} \sum_{i=1}^{N} \frac{1}{c_{i}*V_{com,i}}
\end{equation} 
where $\Delta L$ is the width of the bin used (in our case, 0.5\,dex), $c$ is the  completeness correction factor of the $i$-th galaxy, and $V_{com,i}$ is the co-moving volume of the $i$-th galaxy. The completeness is taken from \cite{backhaus2025} and \cite{Stone2024} for MEGA and SMILES, respectively.
The co-moving volume represents the volume of space an individual galaxy can be detected given the survey's observational depths.
$V_{com,i}$ is calculated using the \cite{Avni1980} method for coherent analysis of independent data-sets: 

\[
V_{com,i} = 
\begin{cases}
\frac{\Omega^{\text{SMILES}}}{4\pi} V_{com,i}^{\text{SMILES}} + \frac{\Omega^{\text{MEGA}}}{4\pi} V_{com,i}^{\text{MEGA}}& \text{if } z_{max,i} \ge z_{max,i}^{\text{MEGA}}  \\
 \frac{\Omega^{\text{SMILES}}}{4\pi} V_{com,i}^{\text{SMILES}} & \text{if } z_{max,i} < z_{max,i}^{\text{MEGA}}
\end{cases}
\]

For each field, $\Omega$ is the solid angle of that field. The co-moving volume of the $i$-th galaxy in a given redshift bin is expressed as $V_{com,i}=V_{z_{max,i}} - V_{z_{min,i}}$ where $z_{min,i}$ is the lower boundary of the considered redshift bin and $z_{max,i}$ selects the smaller value between the upper boundary of the considered redshift bin or the maximum redshift at which a source would be included in the sample given the limiting flux density. The depth between the two surveys is not the same, and therefore an object with a given flux density could  be observed in either one or both samples. For example, sources detected in the MEGA survey are detectable over the whole joint area while the fainter sources detected in SMILES are detectable in SMILES only. 


\subsection{Removing AGNs}

\begin{figure}[b!]
\centering
\includegraphics[width=\linewidth]{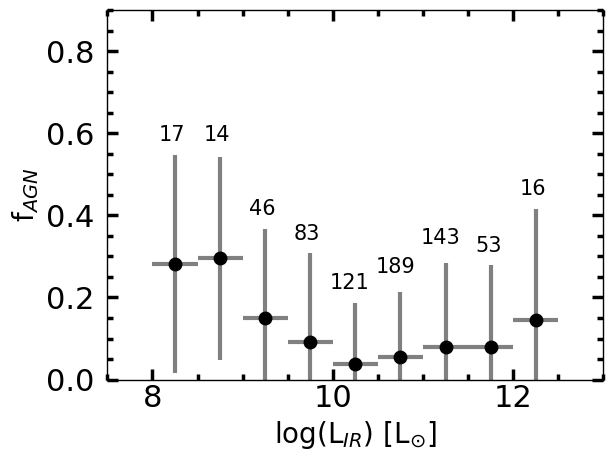}
\caption{Fraction of total $L_{\rm IR}$ attributable to AGN heating ($f_{\rm AGN}=L_{\rm IR}^{\rm AGN}/L_{\rm IR}^{\rm tot}$ in $L_{\rm IR}$ bins for 683 MEGA sources from \citet{Hamblin2025}. $L_{\rm IR}^{\rm tot}$ and $L_{\rm IR}^{\rm AGN}$ were calculated by integrating the best fit {\tt CIGALE} models from $8-1000\,\mu$m. The black points show the average in each bin. The vertical grey lines illustrate the standard deviation, while the horizontal grey lines show the bin width. The total number of sources in each bin is listed above the points. We take the average $f_{\rm AGN}$ as the correction factor to remove AGN heating from each luminosity bin in the final LFs.
\label{fig:AGN frac} }
\end{figure}

We aim to calculate the $L_{\rm IR}$ attributable to star formation only, in order to estimate the amount of a dust obscured star formation at $z>0$. To remove the AGN contribution, we apply a single correction to each luminosity bin. \citet{Hamblin2025} performed spectral energy distribution fitting of MEGA galaxies using {\tt CIGALE} \citep{boquien2019,yang2022} and obtained acceptable fits for 683 MEGA sources ($\chi^2_{\text{red}} \leq 5$). We integrated the best fit model over $8-1000\,\mu$m to calculate the total $L_{\text{IR}}^{\rm tot}$, and we do the same for the AGN model to calculate $L_{\rm IR}^{\rm AGN}$. The AGN fraction for each source is defined as $f_{\rm AGN}=L_{\rm IR}^{\rm AGN}/L_{\rm IR}^{\rm tot}$.
Figure \ref{fig:AGN frac} shows the distribution of $f_{\rm AGN}$ in discrete luminosity bins. We use the average $f_{\rm AGN}$ (black points) as the correction factor for each bin. We opt not to remove sources with AGN heating from the sample entirely as they can still have a significant fraction of $L_{\rm IR}^{\rm tot}$ attributable to star formation. Removing them would artificially lower the LFs and resulting SFRD. We also choose to correct the bin as a whole, rather than individual sources, due to the uncertainties inherent in SED fitting, where good fits depend upon the choice of templates and the availability of high SNR observations covering the optical, near-IR, and mid-IR. Many of our sources lack complete coverage of the mid-IR, due to only emission features being detected (e.g., a PAH feature may fall into a filter, but adjacent filters fail to detect the fainter continuum) or absorption features falling into a given band. This lack of coverage can significantly impact SED fitting of the mid-IR, which is crucial for detecting obscured AGN \citep{Hamblin2025}. Therefore, we choose to rely upon a single conversion factor derived from well-sampled sources, with the implicit assumption that sources with incomplete data have essentially the same SED properties and same volume distribution. The LFs without the AGN contribution removed are shown as the grey points in Figure \ref{fig:LF_methods}. The correction makes the most significant difference at the bright ends. Throughout the remainder of this paper, $L_{\rm IR}$ refers to the AGN-corrected infrared luminosity.

The uncertainties for the final IR LFs includes the Poisson error as well as the standard deviation of $f_{\rm AGN}$ in each bin (grey vertical lines in Figure \ref{fig:AGN frac}), while the monochromatic LF just uses the Poisson error.



\section{Monochromatic Luminosity Functions}
\label{Mono_LF}

\begin{figure*}[t]
\centering
\includegraphics[width=\linewidth]{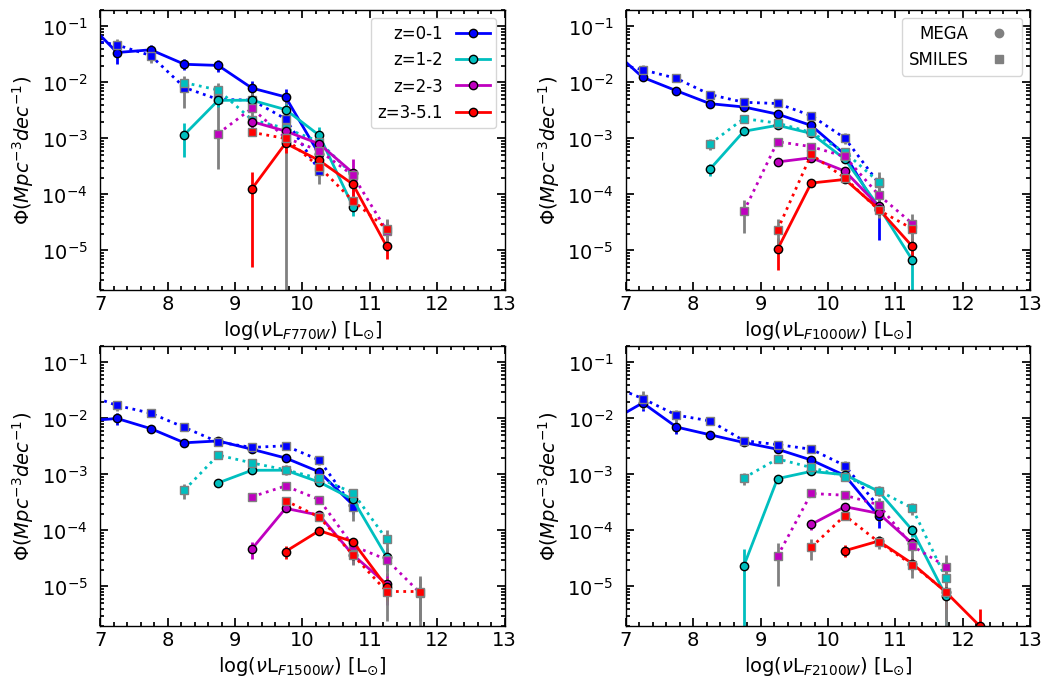}
\caption{Monochromatic LFs for four redshift ranges derived using the F770W, F1000W, F1500W, and F2100W photometry. After calculating $L_\nu$ for a galaxy, we multiplied by the central wavelength in each filter to obtain $\nu L_\nu$. The MEGA observations are shown as points linked with a solid line while the SMILES observations are shown as squares with a dotted line. Both surveys have comparable LFs at each redshift range, with the slight differences likely attributable to cosmic variance.}
\label{fig:filt_LF}
\end{figure*}

We derive the F770W, F1000W, F1500W, and F2100W (corresponding to 7.7, 10, 15, and 21$\mu m$ observed-frame) LFs for the parent SMILES and MEGA samples (Table 1). We calculate the luminosities, $\nu L_\nu$, for each survey individually to test the effects of cosmic variance. We have divided the samples into four redshift bins: $0.0 < z \leq 1$; $1<z\leq2$; $2<z\leq3$; and $3 < z \leq5.1$. The results are shown in Figure \ref{fig:filt_LF}. Generally the SMILES LFs are slightly higher than those of MEGA, which can be due to cosmic variance. Cosmic variance is more pronounced when comparing small-area fields. 

The monochromatic LFs all show a decline toward higher luminosities, indicating that low-luminosity galaxies are much more common than highly luminous systems, as expected. They also reveal an evolution with redshift: all filters indicate that galaxies at earlier cosmic times are, on average, more IR luminous. Moreover, the overall number density of galaxies becomes smaller with increasing lookback time at fixed luminosity. Together, these results hint at a change in characteristic luminosity and abundance. The effects are most pronounced in the two reddest filters, F1500W and F2100W. However, the interpretation of the different evolution seen in the F770W, F1000W versus the F1500W, F2100W is not straightforward. For F770W and F1000W, the rest-frame emission of the galaxies occurs mainly in the near-IR from $z\sim1-5$, and so, for most sources, the photometry is tracing the older stellar population, and the detectability of a galaxy is not strongly evolving. The same is not true in the redder filters. For F1500W, the rest-frame emission covers PAH features at $z\sim1$, which enhance a galaxy's detectability. At $z\sim2-4$, F1500W covers the stellar minimum ($\lambda\sim3-5\,\mu$m), decreasing detectability. Finally, at $z\sim5$, F1500W begins to trace the stellar bump (which peaks at $\lambda\sim1.6\,\mu$m), increasing detectability once again. The bottom right panel of Figure \ref{fig:filt_LF} illustrates this clearly, with the $z\sim2-3$ LF lying significantly below the $z\sim1-2$ LF, while the $z\sim3-5$ LF does not significantly further drop in number density. F2100W follows a similar pattern. There may be fewer galaxies detected around $z\sim1$ as F2100W will cover the rest-frame 9.7\,$\mu$m silicate absorption feature. At $z\sim2$, the PAH complexes begin to fall in the filter, while at $z>3.5$, F2100W covers the stellar minimum. 



\section{Infrared Luminosity Functions }
\label{sec:LF}

\begin{figure*}[t!]
\centering
\includegraphics[width=\linewidth]{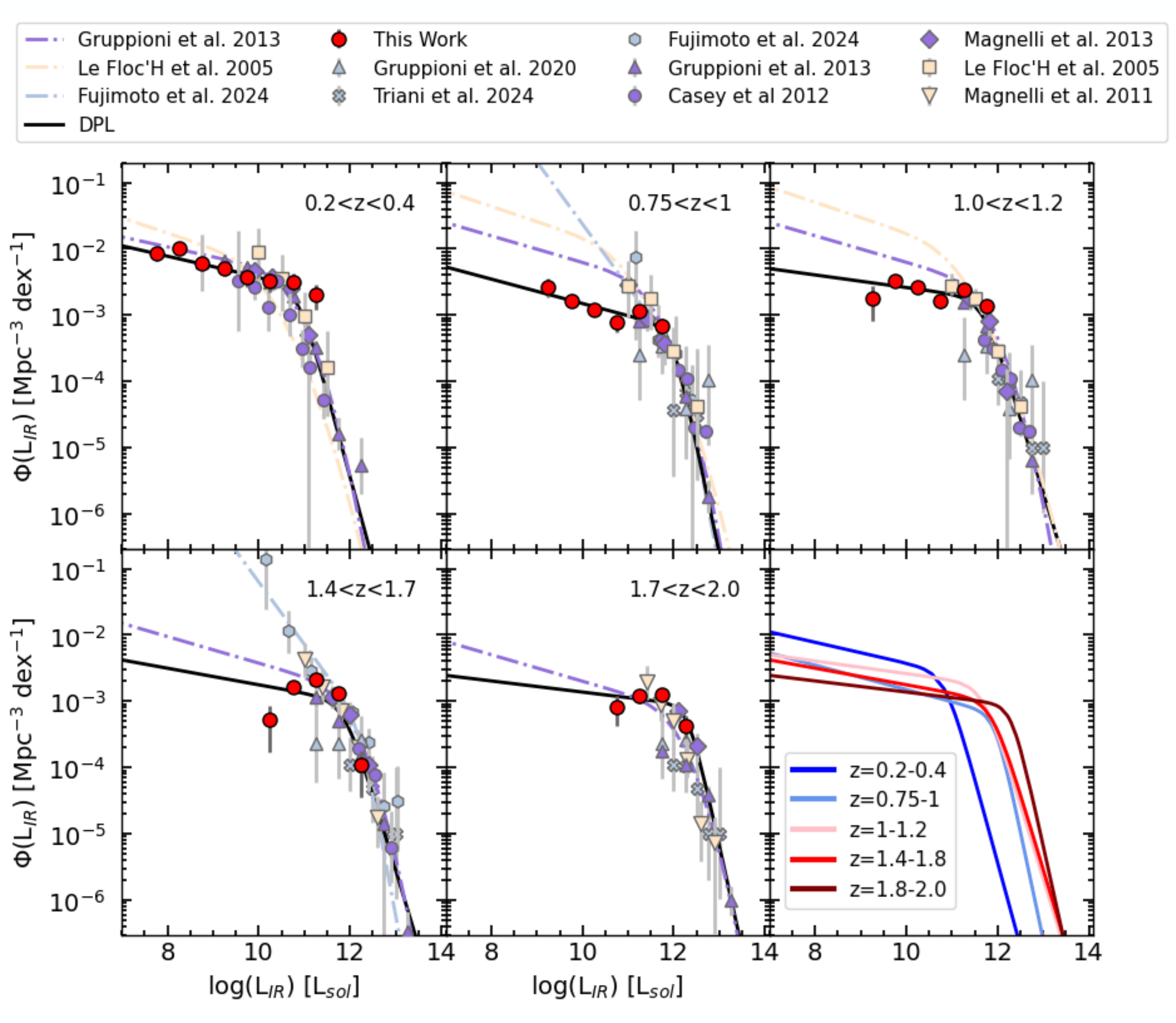}
\caption{IR LFs for five redshift ranges. Previous works using observations from Herschel are shown in purple, Spitzer in pale yellow, and ALMA in blue. The fits to the observations uses a double power-law form whose parameters are given in Section 4.1. The evolution of the LF is shown in the bottom right panel.
\label{fig:LF} }
\end{figure*}

Our IR LFs are shown in Figure \ref{fig:LF}.
Table \ref{tab:1} provides $\Phi(L_{\rm IR}$ in each luminosity and redshift bin, with the number of galaxies in brackets. In Figure \ref{fig:LF}, we also compare to previous works which cover a large range of other telescopes, including those from Herschel \citep{Casey2012,Gruppioni2013,Magnelli2013}, Spitzer \citep{LeFloch2005,magn11}, 
and ALMA \citep{Gruppioni2020,Fujimoto2023,Traina2024}. The literature LFs may not cover the exact same redshift range but are significantly overlapping. Our measurements are generally consistent with previous works at the bright end, indicting that our method for calculating $L_{\rm IR}$ from the 7.7$\mu$m PAH feature is reliable. 


Our data points generally agree with previous measurements, within the uncertainties. Both the SMILES and MEGA photometric catalogs were based solely on mid-IR detections without near-IR priors. Similarly, the \cite{LeFloch2005} and \cite{magn11} LFs were derived from Spitzer 24\,$\mu$m detections, resulting in the good agreement between the Spitzer and MIRI samples. The Herschel samples \citep{Casey2012, Gruppioni2013, Magnelli2013} were all selected based on Spitzer 24\,$\mu$m priors. Due to the large beam sizes of the Herschel/SPIRE instrument, priors are necessary for making secure detections. Hence, the main difference between the Herschel and Spitzer samples is the ability of the former to more robustly measure $L_{\rm IR}$ using far-IR photometry. Therefore, the general agreement between the Herschel, Spitzer, and MIRI samples is expected given the similar wavelength selections. Beyond providing a comparison, including these works also allows us to better constrain the brighter end of the LF (\S4.1), since our own data do not extend beyond $\log L_{\rm IR}=12.2\,[L_{\odot}]$. JWST is hampered by a significantly smaller FOV, making brighter sources much rarer.

Our data points differ significantly at the faint end compared to \cite{Fujimoto2023} in the $0.75<z<1$ and $1.4<z<1.7$ redshift bins. We note for the $1.4<z<1.7$ bin we plot the $1<z<2$ LF from \cite{Fujimoto2023}, though this alone does not account for the difference.
There are several potential causes that can contribute to the discrepancy. First, due to \cite{Fujimoto2023} using lensed galaxies, they must correct for the magnification, and this flux extraction method becomes more uncertain at the faint end. Second, \cite{Fujimoto2023} removes a standard 20\% of the $L_{\rm IR}$ from all sources due to assumed AGN heating, which is much larger than our own AGN correction. This will shift more galaxies into lower luminosity bins. Finally, there is also a difference in how $L_{\rm IR}$ is measured. The \cite{Fujimoto2023} sample fits \citet{draine2001} models, allowing the radiation field to vary as $3<U_{min}<30$. While we do not calculate $L_{\rm IR}$ directly, we rely on the SED fitting in \citet{Shivaei2024}, which also uses \citet{draine2001} models, but constrains the radiation field to $0.1<U_{min}<15$. The choice of radiation field will effect the resulting dust temperature, changing the shape of the far-IR SED. A higher $U_{\rm min}$ corresponds to warmer dust. Based on the SED fitting alone, there is a possibility that \citet{Fujimoto2023} and \citet{Shivaei2024} would derive discrepant $L_{\rm IR}$s for the same source. Further complicating the SED fitting, the \citet{Fujimoto2023} sample lacks any mid-IR observations. They fit to Spitzer/IRAC, ALMA, and Herschel (where available). Although systematics alone may account for the differences in the numbers of fainter sources, sample selection may also play a role. \citet{Fujimoto2023} selects sources based on blind ALMA detections, although the sources do need to have multiwavelength data (in this case, Spitzer/IRAC) for the $L_{\rm IR}$ calculation. \citet{Gruppioni2020} also selects ALMA sources in a similar luminosity range, and their faint end does not agree with \citet{Fujimoto2023}. \citet{Gruppioni2020} uses yet a different SED-fitting technique, this time based on empirical templates rather than models, and most of their sources have Spitzer/MIPS coverage. Given the disagreement between two ALMA-selected samples, the differences in the faint ends of the ALMA samples and our sample are more likely methodological, rather than indicating a strong difference between mid-IR- and submm-detected sources. 




\begin{table*}[t]
  \centering
  \begin{tabular}{|c|c|c|c|c|c|}
  \hline
    log($L_{IR}$) [$L_{\odot}$] & \multicolumn{5}{c|}{log($\Phi(L_{IR})$) [Mpc$^{-3}$ dex$^-1$]} \\
    \hline
     & 0.2<z<0.4 & 0.75<z<1.0 & 1.0<z<1.2 & 1.4<z<1.7 & 1.7<z<2.0\\
    \hline
    7.5-8.0   & -2.08$\pm$0.20 [22] &  --  &  --  &  --  & --\\
    
    8.0-8.5   & -2.00$\pm$0.07 [37] &  --  &  --  &  --  & --\\
    
    8.5-9.0   & -2.22$\pm$0.09 [23] &  --  &  --  &  --  & --\\
    
    9.0-9.5   & -2.29$\pm$0.11 [16] & -2.57$\pm$0.14 [15] &  -2.75$\pm$0.23 [5] &  --  & --\\
    
    9.5-10.0  & -2.43$\pm$0.13 [11] & -2.78$\pm$0.08 [31] & -2.48$\pm$0.08 [37] &  --  & --\\
    
    10.0-10.5 & -2.49$\pm$0.14 [9] & -2.91$\pm$0.09 [22] & -2.59$\pm$0.08 [33] &  -3.29$\pm$0.29 [3] & --\\
    
    10.5-11.0 & -2.50$\pm$0.14 [9] & -3.12$\pm$0.12 [14] & -2.79$\pm$0.09 [21] & -2.78$\pm$0.06 [32] & -3.09$\pm$0.20 [17]\\
    
    11.0-11.5 &  -2.68$\pm$0.17[6]  & -2.93$\pm$0.09[22] & -2.62$\pm$0.08 [39] & -2.67$\pm$0.05 [55] & -2.93$\pm$0.09 [43]\\
    
    11.5-12.0 &  --  &  -3.16$\pm$0.12 [13]  & -2.87$\pm$0.10 [18] &  -2.88$\pm$0.06[34] & -2.89$\pm$0.09 [47]\\
    
    12.0-12.5 &  --  &  --  &  --  &  -3.97$\pm$0.29[3] & -3.36$\pm$0.16 [17] \\
    \hline
  \end{tabular}
  \caption{Infrared luminosity and redshift bins with the number of objects in each bin in brackets.
  \label{tab:1}}
\end{table*}

\subsection{Parameterizing the Luminosity Function}
We fit our data points to create a functional form of the IR LF. As we lack a significant number of high luminosity data points, we augment our data with \cite{Gruppioni2013} to constrain the bright end. We chose to rely solely on \cite{Gruppioni2013} because their measurements are available in every redshift bin, include reported uncertainty, and have shown excellent agreement with the rest of our sample. 
We model the LFs with a double power-law that depends on both redshift and $L_{\rm IR}$:

\begin{equation} \label{Eq:LF}
  \Phi(L,z) = \Phi_\star \left[\left(\frac{L}{L_\star}\right)^{\alpha}+\left(\frac{L}{L_\star}\right)^{\beta}\right]^{-1}
\end{equation}

We perform the fitting using a Markov Chain Monte Carlo (MCMC) approach implemented with the Python package \texttt{emcee} \citep{Foreman2013}, adopting 1000 walkers and 
and 5000 iteration steps. The \texttt{emcee} sampler allows the data points to move within their uncertainties in each iteration, assuming the uncertainties follow a Gaussian profile. By including the \cite{Gruppioni2013} measurements, which tightly constrain the luminosity range around at $L_{\rm IR}
= 10^{11}-10^{13} L_{\odot}$, we do not need to fix the value of $\beta$. Instead, \texttt{emcee} is able to fit all four free parameters ($L_{\star}$, $\Phi_{\star}$, $\alpha$, and $\beta$) simultaneously. The resulting best-fit values and their uncertainties (16th and 84th percentiles) of $\alpha$, $\beta$, $\Phi_\star$, and $L_\star$  are listed in Table \ref{tab:2}. The corresponding double power-law fit is shown as the solid black curves in Figure \ref{fig:LF}.

The faint-end slope of each LF is shallow, with a weighted average value of $\alpha \sim 0.13$. This is significantly lower than the value reported by \cite{Fujimoto2023}, who found a faint-end slope of $\alpha \sim 0.94$ for $z < 2$. As discussed previously, this discrepancy likely arises from 
differences in how $L_{\rm IR}$ was measured. Our faint-end slope is also significantly lower than the commonly adopted value of $\alpha \sim 0.6$ in \cite{magn11} and \cite{Magnelli2013} which comes from the local LF in \citep{sanders2003}. \citep{LeFloch2005} also holds the faint end slope fixed using the same local LF. 
\cite{Gruppioni2013} fit a modified-Schechter function \citep{Saunders1990} to only the $z<0.3$ data and find a faint end slope of $\alpha\sim1.2$. They hold $\alpha$ constant at all other redshift bins. In a double power law model, this would correspond to $\alpha\sim0.2$, in better agreement with our results. 

In contrast, the bright end slope in our analysis is largely constrained by the \cite{Gruppioni2013} measurements, yielding a  weighted average of $\beta \sim 2.60$, in good agreement with \cite{magn11} and \cite{Magnelli2013}. 
\cite{Fujimoto2023} used a double power law and fit the bright end using only the galaxies in their $z=1-2$ bin. They found a bright end slope of $\beta\sim3.72$, which is larger than our slope in every redshift bin. 

\begin{figure}[b]
\centering
\includegraphics[width=\linewidth]{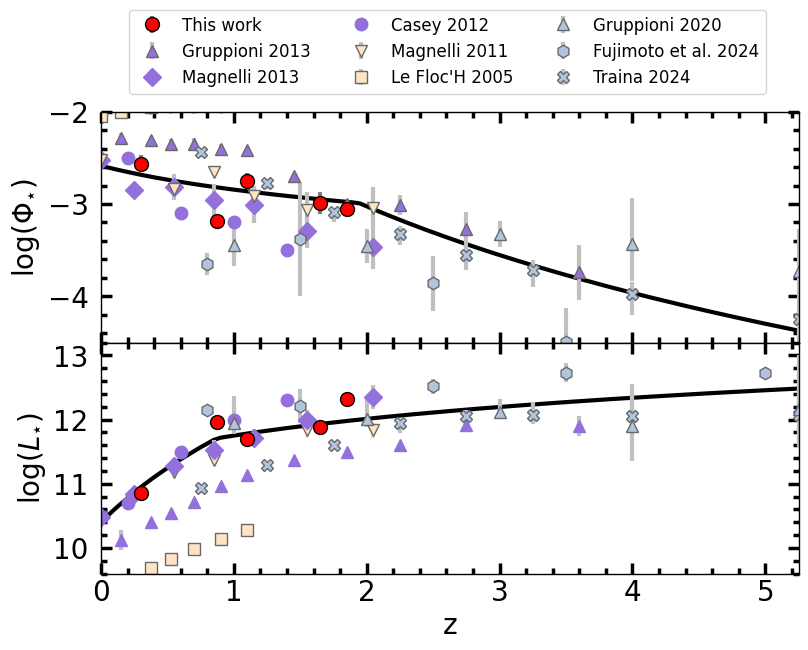}
\caption{Evolution of $\Phi_\star$ and $L_\star$ with same marker notation as Figure \ref{fig:LF}. Our measurements for $\Phi_\star$, $L_\star$ are in agreement with previous works. Our LFs demonstrate a clear evolution in both number density and characteristic luminosity.}
\label{fig:IL_star}
\end{figure}

The parameters $\Phi_\star$ and $L_\star$ can be expressed as: 


\begin{align}
\label{eq:phistar}
\Phi_\star = 
\begin{cases}
  \Phi_{0}(1+z)^{\gamma_1} & \text{if z<z$_{\Phi,\text{turn}}$} \\
  \Phi_{0}(1+z)^{\gamma_2}\times(1+z_{\Phi,\text{turn}})^{\gamma_{1}-\gamma_{2}}& \text{if z>z$_{\Phi,\text{turn}}$}
\end{cases}
\end{align}

and

\begin{align}
\label{eq:lstar}
L_\star = 
\begin{cases}
  L_{0}(1+z)^{\Psi_1} & \text{if z<z$_{L,\text{turn}}$} \\
  L_{0}(1+z)^{\Psi_2}\times(1+z_{L,\text{turn}})^{\Psi_{1}-\Psi_{2}}& \text{if z>z$_{L,\text{turn}}$}
\end{cases}
\end{align}
where $\Phi_0$ and $L_0$ are the characteristic density and luminosity at $z=0$, and $z_{\rm turn}$ represents the redshift where the evolution of $\Phi_\star$ and $L_\star$ change slopes.
We calculate each parameter using the same MCMC method, fitting our values of $\Phi_{\star}$ and $L_{\star}$ together with those from \cite{Gruppioni2013} and \cite{Traina2024} to better constrain the high redshift regime. The resulting best-fit parameters are:
\begin{align}
\log(\Phi_{0})&=-2.59\pm0.05\\
z_{\Phi,{\rm turn}}&=1.94\pm0.08 \\\nonumber
\gamma_{1}&=-0.86\pm0.12 \\ \nonumber 
\gamma_{2}&=-4.22\pm0.32 \\[0.6em] \nonumber 
\log(L_{0})&=10.43\pm0.03 \\\nonumber
z_{L,{\rm turn}}&=0.87\pm0.09 \\\nonumber
\Psi_{1}&=4.81\pm0.46 \\ \nonumber
\Psi_{2}&=1.48\pm0.16
\end{align}
These evolutionary fits, together with measurements from other studies, are presented in Figure \ref{fig:IL_star}.

\begin{table*}[t]
  \centering
  \begin{tabular}{|c|c|c|c|c|}
  \hline
     & $\alpha$ & $\beta$ & $\Phi_{\star}$ & $L_{\star}$ \\
    \hline
    0.2<z<0.4  & 0.16$\pm$0.04 & 2.53$\pm$0.43  &  2.74e-3$\pm$5.73e-4 &  7.30e10$\pm$7.27e9\\
    0.75<z<1.0  & 0.18$\pm$0.04 & 3.27$\pm$0.39  &  6.51e-4$\pm$1.15e-4  &  9.04e11$\pm$1.09e11 \\
    1.0<z<1.2   & 0.10$\pm$0.05 &  2.25$\pm$0.16  &  1.79e-3$\pm$3.54e-4 &  4.98e11$\pm$9.449e10  \\
    1.4<z<1.7   & 0.13$\pm$0.08 & 2.29$\pm$0.12 &  1.03e-3$\pm$2.83e-4 &  7.69e11$\pm$1.61e11  \\
    1.7<z<2.0  & 0.08$\pm$0.09 & 3.12$\pm$0.33 & 8.74e-4$\pm$2.01e-4 &  2.08e12$\pm$3.400e+11  \\
    
    \hline
  \end{tabular}
  \caption{Parameter values of the fitted IR LFs}
  \label{tab:2}
\end{table*}

Our measurements for $\Phi_\star$ are consistent with previous studies. Many previous works found the turning point $z_{\Phi,\text{turn}}\sim1-1.1$, at the beginning of cosmic noon \cite{Gruppioni2013,magn11,Magnelli2013,Traina2024}, rather than our value of $z_{\Phi,{\rm turn}}=1.94\pm0.08$.

While estimates of $L_\star$ are also broadly consistent with earlier work, they tend to lie toward the upper end of the reported range. Our turnover redshift is similar to those found in \cite{magn11} and \cite{Magnelli2013}, but significantly lower than $z_{L,\text{turn}}=2$ and $z_{L,\text{turn}}=3.03$ found in \cite{Gruppioni2013} and \cite{Traina2024}. Having a lower turnover redshift here indicates that the rapid increase in characteristic luminosity does not continue to high redshift, leading to the peak of obscured star formation occurring at relatively lower redshift. This, paired with the higher $z_{\Phi,\text{turn}}$, causes a flattening in the resulting star formation rate density.

\subsection{Discussion: The Significance of the IR Luminosity Function}
Due to the sensitivity of JWST/MIRI and the relatively large-area, multi-band mid-IR surveys MEGA and SMILES, we are now able to directly constrain the faint end slope of the IR LF with measurements beyond the local universe. Previous Spitzer and Herschel results (shown in Figure \ref{fig:LF}) had to rely largely on extrapolations from a single, well-sampled low redshift bin, or the local IR LF \citep{sanders2003}, due to the sensitivity limits of the telescopes. Even when our faint end slopes are similar \citep{LeFloch2005,Gruppioni2013}, the literature LFs have a much higher normalization, predicting more faint sources. Perhaps more interestingly, we find a slight evolution in $\alpha$, such that the slope becomes even flatter with increasing redshift. This is most apparent when examining the $z=1-1.2$ bin, where we are able to sample well below the knee of the LF. Suddenly, $\alpha$ drops from 0.18 to 0.10 at $z=1$.

This flattening has two possible contributions. First, at lower masses, there may simply be fewer galaxies with significant amounts of dust. It has been shown that (U)LIRGs ($L_{\rm IR}\ge 10^{11}\,L_\odot$) have strongly evolving gas fractions with increasing lookback time \citep{magdis2012,scoville2017}, leading to higher dust masses than their local counterparts \citep{kirkpatrick2017a}. Environment may also play a role in this evolution, as dust-rich galaxies may be more prone to lie in dense environments \citep{aravena2010,smolcic2017,crespo2021} or be interacting \citep{ren2025}. It is possible that below some characteristic mass, the gas fraction does not evolve as strongly, either due to environment, halo mass, or some other factor, leading to less dust within the galaxy and a decreased detectability in the IR.

Second, within lower mass galaxies, dust-obscured star formation may not be the dominant mode, as it is in more massive galaxies. Indeed, this is supported by the examination of the near-IR and mid-IR SEDs in \citet{Kirkpatrick2023}. That work sorted MIRI-detected galaxies into bins of $L_{\rm IR}$ to create average empirical SEDs and found that the ratio of near-IR to mid-IR emission evolved strongly with decreasing $L_{\rm IR}$. Below $L_{\rm IR}=10^{10}\,L_\odot$, the stellar near-IR emission was much brighter than the PAH emission, indicating a declining dust mass-to-stellar mass ratio as galaxies decrease in brightness. Unlike their brighter counterparts, unobscured star formation seems to remain the dominant mode in less massive galaxies at cosmic noon.

Comparison with the UV LF, which traces the unobscured star formation, can be illuminating.
Many works note the faint end UV LF slope being steep ($\alpha<-1.5$) at $z < 3$ \citep[e.g.][]{Oesch2010,Alavi2016}. 
Previous works have also generally found that the faint end slope continues to steepen as redshift increases \citep[e.g.][]{Oesch2010,Alavi2016,Parsa2016,Bouwens2021,Sun2026}. This behavior supports the interpretation that less massive galaxies, throughout cosmic time, never seem to be dominated by obscured star formation. 
Both the UV and IR LF share a similar evolution in characteristic density, with a general decrease with increasing redshift \citep[e.g.][]{Alavi2016,Parsa2016,Bouwens2021}. Interestingly, while both functions show an increasing characteristic luminosity up to cosmic noon, we find continue to see an increase at earlier times, while $L_\star^{\rm UV}$ appears to start decreasing \citep[e.g.][]{Alavi2016,Parsa2016,Bouwens2021}. It would be unwise to overinterpret this discrepancy, however. The faint end of UV luminosity function can be sampled out to high redshifts, even before JWST, thanks to the sensitivity of the Hubble Space Telescope. The same is not true for the IR luminosity function, which relies on measuring dust-rich galaxies with Herschel or submm observatories beyond $z>3$, leading to a dearth of faint sources. The high redshift knee of the IR LF is consequently much more difficult to reliably measure, and the true evolutionary behavior may not yet be known.

\section{Cosmic SFR Density Evolution}\label{SFRD}
\begin{figure}[b!]
\centering
\includegraphics[width=\linewidth]{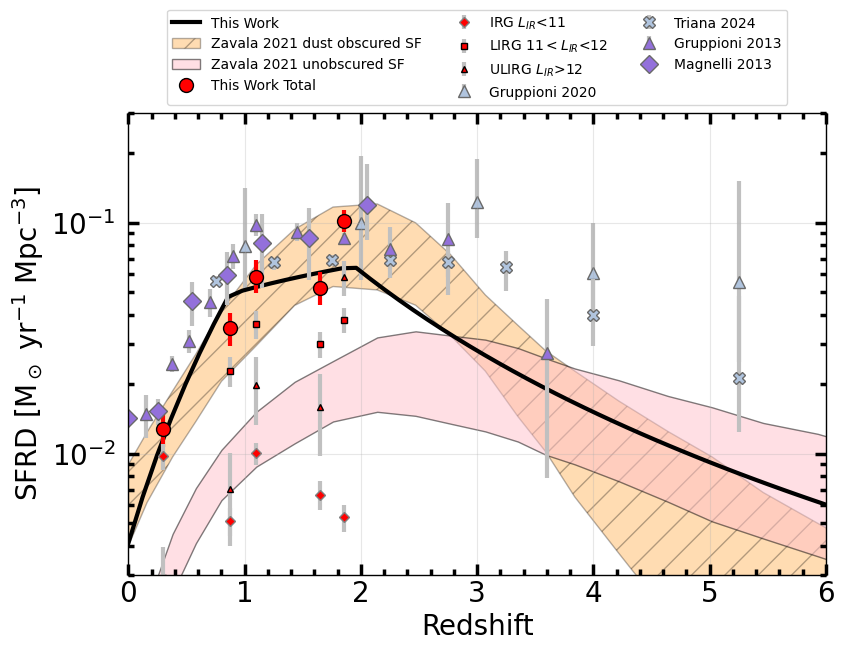}
\caption{Our SFRD (black line), which is calculated using the evolutionary trends of $\Phi_\star$ and $L_\star$, with the individual red points showing $\rho_{\rm SFR}$ calculated from integrating the LFs in Figure \ref{fig:LF}. Beyond $z\sim2$, our SFRD is smooth as it is extrapolated from the evolution of $\Phi_\star$ and $L_\star$ in Figure \ref{fig:IL_star}. Our SFRD from $z=0-2$ aligns well with previous works. We break our SFRD down into the contribution from ULIRGS ($L_{\rm IR}\ge10^{12}\,L_\odot$), LIRGS ($10^{11}\le L_{\rm IR}<10^{12}\,L_\odot$), and IRGS ($L_{\rm IR}<10^{11}\,L_\odot$) and find that LIRGS are the dominant contributor at most redshifts. 
\label{fig:pSFR} }
\end{figure}

IR emission arises mainly from dust heated by stars, and $L_{\rm IR}$ can be converted to a galaxy's average SFR over the past 100 million years. By considering the $L_{\rm IR}$ functions, we can calculate the SFRD of the universe over cosmic time, where SFRD represents the total amount of star formation in a given cosmic volume.
We first compute the $L_{\rm IR}$ density ($\rho_{\rm IR}$) in each redshift bin by integrating our best-fit IR LF according to: 

\begin{equation} \label{Eq:rIR}
  \rho(IR) = \int_{8}^{14}\Phi(L,z)L\ dlog(L)
\end{equation}
where the integration limits run from $L_{\rm IR}=10^8-10^{14}\,L_\odot$, which is comparable to the limits adopted in previous works \citep[e.g.][]{Traina2024,Ling2024,Gruppioni2020}. Below this limit, galaxies are too faint to meaningfully contribute to the SFRD, and above this limit, galaxies are too rare to meaningfully contribute. The 1$\sigma$ uncertainties are derived from the 16th and 84th percentiles of the posterior probability distributions obtained from the MCMC fitting.
Since we have corrected for the AGN contribution when calculating the IR LF, $\rho_{IR}$ can be directly used as a proxy for the SFRD ($\rho_{SFR}$) via the relation of \cite{kenn98} for a \cite{chab03} IMF: $\rho_{SFR} = 1.09\times10^{-10} \rho_{\rm IR}$, where the SFR has units of $M_\odot/{\rm yr}$ and $L_{\rm IR}$ has units of $L_\odot$. Figure \ref{fig:pSFR} shows $\rho_{\rm SFR}$ as the red points, which come directly from integrating the LFs.  
Our $\rho_{SFR}$ values are generally in good agreement with previous measurements based on ALMA, Herschel, and Spitzer, within the uncertainties 

\citep{Magnelli2013,Gruppioni2013,Gruppioni2020,Traina2024}.

We calculate SFRD over cosmic time (black line) by assuming a standard form for the IF LF, fixing $\alpha$ and $\beta$ to our average values from all five redshift bins. Then, we use our evolution of $\Phi_\star$ and $L_\star$ (Figure \ref{fig:IL_star}) to calculate the LF at discrete redshifts. We remind the reader that beyond $z\sim2$, our $\Phi_\star$ and $L_\star$ models were constrained using literature measurements from Herschel and ALMA sources. 
For comparison, we show the dust-obscured and unobscured SFRD based on ALMA observations from \cite{Zavala2021}, and our values agree up to $z=2$. In that work, the authors assume a dust emission model and a double-power law form for the LF. They then use backward modeling of the number counts of galaxies detected from $1-3\,$mm to derive the LF parameters and evolution of $\Phi_\star$ and $L_\star$. The bright end slope of the LF is held fixed to be $\beta=3.0$, while the faint end slope is derived to be $\alpha=0.42$. Notably, the \citet{Zavala2021} faint end slope is much steeper than ours, and they hold it constant for all redshifts. Their parameters for $L_\star$ evolution (Equations \ref{eq:phistar} and \ref{eq:lstar}) also vary significantly, in that they adopt $\Psi_1=2.8$ (we fit $\Psi_1=4.81$) and $\Psi_2=1.0$ (similar to our $\Psi_2=1.48$). Crucially, the \citet{Zavala2021} density evolution is much different than ours. They fix $z_{\Phi,{\rm turn}}=2.0$ (similar to our fitted $z_{\Phi,{\rm turn}}=1.94$), with no evolution, $\gamma_{1}=0$, below this redshift. They fit a quite steep evolution above $z=2$, with $\gamma_2=-6.5$ (we fit $\gamma_{2}=-4.22$). This steep evolution is evident at $z>3$ in Figure \ref{fig:pSFR} and explains the discrepancy with our work at the higher redshift end of the SFRD. Our SFRD declines more steeply at $z>3$ than the \citet{Gruppioni2013} and \citet{Traina2024} data points. 
This arises mainly from a difference in $z_{L,{\rm turn}}$ and $z_{\Phi,{\rm turn}}$ between our works, as \citet{Traina2024} found $z_{L,{\rm turn}}=3.03$ and $z_{\Phi,{\rm turn}}=0.89$.

The true evolution of the dust-obscured SFRD at $z>3$ can only be robustly constrained by better modeling of the IR LF, particularly with more faint sources to constrain the knee and faint end slope. Faint sources are especially required as the number of luminous, or massive, dusty galaxies is expected to decrease with increasing redshift. Observing large numbers of main sequence dusty galaxies is beyond the capabilities of current observatories. JWST is not probing dust emission much beyond $z=3$, even in the longest wavelength filters. Therefore, JWST cannot be used to directly or reliably measure $L_{\rm IR}$ at these redshifts. Submm observatories are not capable of building up statistical samples of faint sources in a reasonable amount of observing time, and most of the IR SED would still need to be extrapolated due to lack of complementary data sets in this range. The only way to directly measure the obscured star formation rate is through a new, sensitive far-IR observatory, such as PRIMA \citep{glenn2025}. 



\begin{table}[b]
  \centering
  \begin{tabular}{|c|c|c|c|c|}
  \hline
    & \multicolumn{4}{c|}{log($\rho_{SFR}$)}\\
\hline

     z & Total & IRG & LIRG & ULIRG \\
    \hline
    \small{0.2<z<0.4}  & \small{-1.89$\pm$0.08} & \small{-2.01$\pm$0.06}  &  \small{-2.54$\pm$0.16} & \small{-3.99$\pm$0.80}\\
\small{0.75<z<1.0}  & \small{-1.45$\pm$0.07} & \small{-2.29$\pm$0.06}  &  \small{-1.64$\pm$0.06}  & \small{-2.15$\pm$0.19}\\
\small{1.0<z<1.2}   & \small{-1.23$\pm$0.08} &  \small{-2.00$\pm$0.05}  & \small{-1.43$\pm$0.06}   & \small{-1.70$\pm$0.14}\\
\small{1.4<z<1.7}   & \small{-1.28$\pm$0.07} & \small{-2.18$\pm$0.06} &  \small{-1.53$\pm$0.06}  & \small{-1.80$\pm$0.17}\\
\small{1.7<z<2.0}  & \small{-0.99$\pm$0.05} & \small{-2.27$\pm$0.06} & \small{-1.42$\pm$0.05} & \small{-1.24$\pm$0.07}\\
    \hline
  \end{tabular}
  \caption{$\rho_{\rm SFR}$ points obtained by integrating the LF in our five redshift bins.}
  \label{tab:4}
\end{table}


We decompose the SFRD into contributions from infrared galaxies (IRGs; $L_{\rm IR}<10^{11}\,L_\odot$), luminous infrared galaxies (LIRGs; $10^{11}\le L_{\rm IR}<10^{12}\,L_\odot$), and ultraluminous infrared galaxies (ULIRGs; $L_{\rm IR}\ge10^{12}\,L_\odot$). We find that IRGs dominate the SFRD at $z<0.6$, which is evidence of cosmic downsizing. More luminous galaxies run through their gas supply more quickly due to their intense SFRs, leaving less luminous galaxies as the dominate sites of star formation today. LIRGs become the dominate sites of star formation over $0.6<z<1.75$. Near the peak of cosmic noon, $z\sim2$, ULIRGS contribute the most to the SFRD. These trends are broadly consistent with the results of \cite{Murphy2011}, which found that IRGs dominate at $z<0.7$ and ULIRGs dominate at $z=2$. Notably, \citet{Murphy2011} found that IRGs and LIRGs to contribute equally at $z>0.7$. 
This discrepancy likely arises from differences in the faint end slope of the LF. Due to a lack of faint Spitzer 24\,$\mu$m sources, \citet{Murphy2011} adopted the same functional form of the LF as \cite{Magnelli2009,magn11}, which uses $\alpha = -0.6$ for the faint end slope. This is significantly steeper than our measurement. A steeper faint-end slope naturally yields a higher IRG contribution to the total SFRD. By now directly measuring the faint end of the LF, we show that IRGs are relatively unimportant sites of cosmic stellar mass assembly at $z=0.5-2.0$.  

\section{Summary}\label{Summary}
We combine JWST/MIRI observations from the MEGA and SMILES survey in order to build IR LFs.
From this parent sample, we select 634 galaxies spanning the redshift range $0.2<z<2$, where the 7.7$\mu$m PAH feature falls within the three redder MIRI bands (F1000W, F1500W, and F2100W). We use the photometric flux to estimate $L_{7.7PAH}$ and covert to $L_{\rm IR}$. We calculate monochromatic LFs and the IR LF in five redshift bands, where MIRI observations are tracing the rest-frame dust emission. We use the IR LF to measure the evolution of $\Phi_\star$ and $L_\star$. Finally, we calculate $\rho_{\rm IR}$ and convert to $\rho_{\rm SFR}$ to measure the dust-obscured SRFD over cosmic time. 
Our main results can be summarized as follows:

\begin{itemize}
    \item Our method of measuring $L_{\rm IR}$ from the 7.7$\mu$m PAH feature is reliable from $z=0-2$ based on comparison with independent measurements in previous works sampling the full SED from Spitzer, Herschel, and ALMA observations. 
    The correlation between the 7.7$\mu$m PAH feature and $L_{\rm IR}$ provides a powerful tool for estimating $L_{\rm IR}$ from JWST observations alone.

    \item Our redder monochromatic LFs (F1500W and F2100W) illustrate a significant evolution with redshift which can be attributed to MIRI observations tracing significantly different regimes of the rest-frame SED. Beyond $z\sim2$ ($z\sim3$), F1500W (F2100W) is no longer tracing dust emission. MIRI observations cannot be used past these redshifts to probe dust emission or infer $L_{\rm IR}$. 

    \item Our LFs extend to fainter luminosities than studies using the previous generation of IR telescopes, allowing us to directly constrain the faint end slope. This improved depth reveals a shallow slope ($\alpha<0.2$), in stark contrast to previous assumptions and extrapolations based on the local LF. Unlike their luminous counterparts, faint galaxies with an abundance of dust are either rarer than previously thought, or dust-obscured star formation is not the dominant mode in less massive galaxies.
    
    \item We find no significant change in the shape of the LF in the faint- and bright- end slopes ($\alpha$, $\beta$) over the redshift range $0.2<z<2$, while both $L_\star$ and $\Phi_\star$ evolve strongly with redshift. When paired with the observed evolution in the SFRD, this indicates that cosmic star formation  is primarily driven by changes in the characteristic luminosity and number density of IR-emitting galaxies, rather than by a redshift-dependent reshaping of the LF.

    \item We derive the obscured SFRD and find that it rises over $0<z<1$ and then flattens between $1<z<2$. At low redshift ($z<0.6$), normal IRGs ($L_{\rm IR}<10^{11}\,L_\odot$) dominate the SFRD, while at intermediate redshifts ($0.6<z<1.75$), LIRGs ($10^{11}\,L_\odot\le L_{\rm IR}<10^{12}\,L_\odot$) become the primary contributors. Near $z \sim 2$, LIRGs and ULIRGs ($L_{\rm IR}\ge10^{12}\,L_\odot$) contribute nearly equally to the total SFRD.

    \item Beyond $z\sim2$, the obscured SFRD remains uncertain, with conflicting results in the literature depending sensitively on how $\Phi_\star$ and $L_\star$ evolution was modeled. Due to the limitations of current facilities, particularly the fact that JWST does not trace dust emission beyond $z\sim3$, concretely pinning down the high redshift obscured SFRD will require a new, sensitive far-IR facility capable of observing statistical populations.

\end{itemize}
JWST/MIRI observations provide a critical means of constraining the faint end of the infrared LF, thereby revealing how dust evolves and how it affects measurements of star formation. These constraints on the faint end will continue to improve with additional JWST/MIRI observations and advances in SED modeling, which will enable us to probe progressively fainter galaxies. 

\vspace{1.5mm}

\section{Acknowledgments}
We would like to thank Dr. Alexandra Pope for her helpful insight and conversations to improve this work. We would also like to thank Dr. Stacey Alberts for her significant effort in producing the publicly available SMILES data set.
This work was financially supported by JWST-GO-03794.001, which also provided the MEGA dataset. K.H. acknowledges financial support from NASA/FINESST 21-ASTRO21-0087.
The MEGA and SMILES (JWST-GTO-1207) JWST data presented in this paper were obtained from
the Mikulski Archive for Space Telescopes (MAST) at the Space Telescope Science Institute. 
\software{\texttt{AstroPy} \citep{astropy2013}, \texttt{Matplotlib} \citep{hunter2007}, \texttt{NumPy} \citep{vanderwalt2011}, \texttt{SciPy} \citep{jones2001}, \texttt{eazy-py} \citep{bram08}, ASTRORMS, Photutils \citep{Bradley2023};JWST Calibration Pipeline \citep{bushouse_howard_2022_7429939}; SEXTRACTOR \citep{Bertin96};SEP \citep{Barbary2016}}

\restartappendixnumbering
\appendix

\section{Different IR Conversion Method}

Table \ref{tab:2} presents the LF for the AGN corrected combined \cite{Houck2007} and \cite{Shivaei2024} conversion. Here we present the alternative LF values shown in Figure \ref{fig:L_meth_dif} in Table \ref{tab:method}. These alternatives include an uncorrected AGN version of the combined method, an AGN corrected  \cite{Houck2007} conversion, and an AGN corrected \cite{Shivaei2024} conversion.

\begin{table*}[h]
  \centering
  \begin{tabular}{|c|c|c|c|c|c|}
  \hline
    log($L_{IR}$) [$L_{\odot}$] & \multicolumn{5}{c|}{log($\Phi(L_{IR})$) [Mpc$^{-3}$ dex$^-1$]} \\
    \hline
     AGN uncorrected & 0.2<z<0.4 & 0.75<z<1.0 & 1.0<z<1.2 & 1.4<z<1.7 & 1.7<z<2.0\\
    \hline
    7.5-8.0   & -1.93$\pm$0.10[22] &  --  &  --  &  --  & --\\
    
    8.0-8.5   & -1.85$\pm$0.07[37] &  --  &  --  &  --  & --\\
    
    8.5-9.0   & -2.06$\pm$0.09[23] &  --  &  --  &  --  & --\\
    
    9.0-9.5   & -2.22$\pm$0.11[16] & -2.50$\pm$0.14[15] &  -2.68$\pm$0.23[5] &  --  & --\\
    
    9.5-10.0  & -2.38$\pm$0.13[11] & -2.74$\pm$0.08[31] & -2.43$\pm$0.08[37] &  --  & --\\
    
    10.0-10.5 & -2.35$\pm$0.13[9] & -2.89$\pm$0.09[22] & -2.57$\pm$0.08[33] &  -3.25$\pm$0.29[3] & --\\
    
    10.5-11.0 & -2.35$\pm$0.13[9] & -2.86$\pm$0.09[14] & -2.63$\pm$0.08[21] & -2.76$\pm$0.08[32] & -3.06$\pm$0.14[17]\\
    
    11.0-11.5 & -3.12$\pm$0.31[6]  & -3.03$\pm$0.11[22] & -2.60$\pm$0.08[39] & -2.64$\pm$0.06[55] & -2.89$\pm$0.07[43]\\
    
    11.5-12.0 &  --  &  -3.13$\pm$0.12[13]  & -3.89$\pm$0.31[18] &  -2.85$\pm$0.07[34] & -2.86$\pm$0.06[47]\\
    
    12.0-12.5 &  --  &  --  &  --  &  -3.90$\pm$0.25[3] & -3.30$\pm$0.11[17]\\
    \hline
    \hline
    Houck & 0.2<z<0.4 & 0.75<z<1.0 & 1.0<z<1.2 & 1.4<z<1.7 & 1.7<z<2.0\\
    \hline
    7.5-8.0   & -2.00$\pm$0.07[37] &  --  &  --  &  --  & --\\
    
    8.0-8.5   &-2.05$\pm$0.08 [33] &  --  &  --  &  --  & --\\
    
    8.5-9.0   & -2.32$\pm$0.10[18] &  -2.55$\pm$0.13[17]  &  -2.66$\pm$0.20[7]  &  --  & --\\
    
    9.0-9.5   & -2.38$\pm$0.12[13] & -2.81$\pm$0.08[31] &  -2.46$\pm$0.07[45] &  --  & --\\
    
    9.5-10.0  & -2.51$\pm$0.14[9] & -2.88$\pm$0.09[25] & -2.59$\pm$0.07[35] &  -2.83$\pm$0.13[17]  & -3.39$\pm$0.26[4]\\
    
    10.0-10.5 & -2.40$\pm$0.13[11] & -2.91$\pm$0.09[22] & -2.71$\pm$0.09[25] &  -2.71$\pm$0.06[48] & -2.97$\pm$0.07[36]\\
    
    10.5-11.0 & -2.85$\pm$0.21[4] & -2.98$\pm$0.10[19] & -2.75$\pm$0.09[23] & -2.82$\pm$0.07[38] & -2.96$\pm$0.07[39]\\
    
    11.0-11.5 & --  & -3.43$\pm$0.16[7] & -2.92$\pm$0.11[16] & -3.16$\pm$0.10[18] & -3.06$\pm$0.08[32]\\
    
    11.5-12.0 &  --  &  --  & -3.82$\pm$0.31[2] &  -3.72$\pm$0.19[5] & -3.44$\pm$0.12[13]\\

    \hline
    \hline
    Shivaei & 0.2<z<0.4 & 0.75<z<1.0 & 1.0<z<1.2 & 1.4<z<1.7 & 1.7<z<2.0\\
    \hline
    7.5-8.0   & -2.08$\pm$0.10 [22] &  --  &  --  &  --  & --\\
    
    8.0-8.5   & -2.00$\pm$0.07 [37] &  --  &  --  &  --  & --\\
    
    8.5-9.0   & -2.22$\pm$0.09 [23] &  --  &  --  &  --  & --\\
    
    9.0-9.5   & -2.29$\pm$0.11 [16] & -2.55$\pm$0.13 [15] &  -2.67$\pm$0.21 [5] &  --  & --\\
    
    9.5-10.0  & -2.43$\pm$0.13 [11] & -2.78$\pm$0.08 [31] & -2.48$\pm$0.08 [37] &  --  & --\\
    
    10.0-10.5 & -2.49$\pm$0.14 [9] & -2.91$\pm$0.09 [22] & -2.59$\pm$0.08 [33] &  -3.18$\pm$0.25 [3] & --\\
    
    10.5-11.0 & -2.55$\pm$0.15 [8] & -3.11$\pm$0.12 [14] & -2.79$\pm$0.09 [21] & -2.76$\pm$0.08 [32] & -3.07$\pm$0.14 [18]\\
    
    11.0-11.5 & -2.68$\pm$0.18 [6]  & -2.99$\pm$0.10 [19] & -2.76$\pm$0.09 [23] & -2.79$\pm$0.07 [42] & -3.07$\pm$0.08 [31]\\
    
    11.5-12.0 &  --  &  -3.07$\pm$0.11 [16]  & -2.89$\pm$0.11 [17] &  -2.95$\pm$0.08 [29] & -3.03$\pm$0.07 [34]\\
    
    12.0-12.5 &  --  &  -3.67$\pm$0.22 [4]  &  -2.95$\pm$0.11[ 15]  &  -3.30$\pm$0.12 [14] & -3.14$\pm$0.08 [28] \\

    12.5-13.0 &  --  &  --  &  -3.82$\pm$0.31 [2]  &  -3.67$\pm$0.18 [6] & -3.45$\pm$0.12 [14] \\
    \hline
    
  \end{tabular}
  \caption{Infrared luminosity and redshift bins with the LF, $\log(\Phi)$, in units of $[{\rm Mpc}^{-3} {\rm dex}^{-1}]$ with the number of objects in each bin in brackets for the uncorrected combined conversion and different AGN corrected conversion methods.
  \label{tab:method}}
\end{table*}

\bibliography{lib}{}
\end{CJK}

\end{document}